\documentclass[epj]{svjour}
\usepackage{graphics}

\usepackage{tipa}
\usepackage{amssymb}
\usepackage{pifont}
\usepackage{graphicx,subfigure,amsmath}
\usepackage{subfigure}
\usepackage{color}

\usepackage{txfonts}

\usepackage{longtable}
\usepackage{dcolumn}
\usepackage{booktabs}

\usepackage{threeparttable}

\begin{document}
\title{Energy conditions and entropy density of the universe}
\author{Wen-Fei Liu\inst{1,2} \and Jing Niu\inst{2} \and Juan Li\inst{1} \and Tong-Jie Zhang\inst{2,1}
\thanks{\emph{e-mail:} tjzhang@bnu.edu.cn}%
}                     
%
%
\institute{College of Physics and Electronic Engineering, Qilu Normal University, Jinan 250200, China \and Department of Astronomy, Beijing Normal University, Beijing100875, China}
\date{Received: date / Revised version: date}
%
\abstract{
In the standard Friedmann-Lema\^{i}tre-Robertson-Walker (FLRW)
cosmological model, the energy conditions provides model-independent
bounds on the behavior of the distance modulus. However, this method
can not provide us the detailed information about the violation
between the energy conditions and the observation. In this paper, we
present an extended analysis of the energy conditions based upon
the entropy density of the universe. On the one hand, we find that
these conditions imply that entropy density $s$ depends on Hubble
parameter $H(z)$. On the other hand, we compare the theoretical
entropy density from the conservation law of energy-momentum tensor
with that from the energy conditions using the observational Hubble
parameter. When we consider a FLRW universe, according to the theoretical prediction, OHD, thermodynamics and several independent cosmological probes, show that the dominant energy condition is fitter than other energy conditions.
\PACS{
      {energy conditions}, {entropy density}, {OHD}, {thermodynamics}.
     } 
} 
\maketitle
\section{Introduction}
\label{sec:intro}

Over the past decade many pieces of evidence for an accelerated expansion of
the universe have been found with several independent cosmological
probes, such as the supernova (SN) Ia observations~\cite{paper:04,paper:05,paper:06,paper:07,paper:08,paper:29}, cosmic
microwave background (CMB)~\cite{paper:09,paper:10,paper:11}, baryon acoustic oscillations
(BAO)~\cite{paper:12,paper:13,paper:14}, integrated Sachs-Wolfe effect~\cite{paper:15,paper:16,paper:17}, galaxy
clusters~\cite{paper:18,paper:19,paper:20,paper:21} and strong gravitational lensing~\cite{paper:22}. There are
various attempts to explain the acceleration, from dark energy to
modified gravity. Combined analysis of the above cosmological
observations support that an approximately 26\% of cold dark matter
(CDM) and the other part 74\% dominated by an unknown exotic
component with negative pressure-driving the current acceleration.
To study the physical properties that hold for a variety of
matter sources, Hawking and Ellis found the so-called energy
condition[1-3,36], which are invoked in General Relativity to
restrict general energy-momentum tensors. Because these conditions
do not require a specific equation of state of the matter in the universe, they provide very simple and model-independent bounds on
the behavior of the energy density, pressure, and lookback time.
Therefore, the energy conditions are one of many approaches to
understand the evolution of the universe.

However, what can we learn from the energy conditions? In 1997,
Visser~\cite{paper:23} found that current observations indicate that the strong energy condition is violated at some time between the epoch of galaxy formation and the present. This violation implies that no
possible combination of normal matter is capable of fitting the
observational data. In Ref.~\cite{paper:24,paper:25,paper:28}, the conditions are further investigated and found that all the energy conditions seem to be violated at lower redshift. However, this approach has also its limitations. For example, it neither provides us the detailed information about the acceleration expansion, nor the reason for the confrontation between the energy conditions and the observation.

In 1934, Tolman studied the principle of entropy increase on a
periodic sequence of closed Friedmann-Robertson-Walker universes
(Tolman 1934; North 1965). He found that a steady increase in
entropy leads to the growth of radiation pressure. As previous works
have stated,  energy conditions are a series of inequalities between
energy density $\rho$ and pressure $p$. Indeed, it is a challenge to
determine the evolution of energy density and pressure. From a
theoretical point of view, it is not only important to clarify the
relationship between energy conditions and entropy, but, whether the
energy conditions are appropriate to describe the acceleration. This
problem motivates us to study the influence of energy conditions on
the entropy density of the universe and the deceleration parameter
$q(z)$.

This paper is organized as follows. In Sec.2, the energy
conditions are briefly introduced. Next, in Sec.3 we
primarily discuss the detailed evolution of entropy density on the
strength of energy conditions and laws of thermodynamics,
respectively; and then we give the behavior of the first-order
derivative and second-order derivative of entropy density. Moreover,
we compare the behavior of entropy density with Hubble parameter
$H(t)$. In Sec.4, we present the influence of energy conditions
on the deceleration parameter $q(z)$. The conclusions and discussion
are given in Sec.5.

\section{Energy conditions}
\label{sec:theory}

Within the framework of the standard Friedmann-Lema\^{i}tre\- -Robertson-Walker (FLRW) model, the energy-momentum tensor for the perfect fluid can be described by
\begin{equation}
        T_{\mu\nu}=(\rho+p)u_{\mu}u_{\nu}+pg_{\mu\nu}.
        \label{eq:1}
\end{equation}
The total energy density $\rho$ and pressure $p$ of the cosmological
fluid as a function of scale factor $a(t)$ are respectively given by
\begin{equation}
\rho=\frac{3}{8\pi
G}[\frac{\dot{a}^{2}}{a^{2}}+\frac{kc^{2}}{a^{2}}],
\label{eq:2}
\end{equation}

\begin{equation}
p=-\frac{c^{2}}{8\pi G}[2\frac{\ddot{a}}{a}+\frac{\dot{a}^{2}}{a^{2}}+\frac{kc^{2}}{a^{2}}],
\label{eq:3}
\end{equation}
where $k$ denotes the spatial curvature constant with $k=+1$, 0 and
$-1$ corresponding to a closed, flat, and open universe,
respectively. When we consider an FLRW universe, the null, weak,
strong, and dominant energy condition can be expressed as the
following forms (e.g.~\cite{paper:01,paper:02,paper:03,paper:38}):
\begin{eqnarray}
\mbox{\bf NEC} &&
                \Longleftrightarrow  \quad\, \rho + p \geq 0 \;,  \nonumber\\
\mbox{\bf WEC} &&
                \Longleftrightarrow  \quad\, \rho \geq 0
\qquad\, \quad\, \mbox{and}  \quad\, \rho + p \geq 0 \;,  \nonumber\\
\mbox{\bf SEC} &&
                \Longleftrightarrow  \quad\, \rho + 3p \geq 0
\quad\,  \mbox{and} \quad\, \rho + p \geq 0 \;, \nonumber\\
\mbox{\bf DEC} &&
               \Longleftrightarrow  \quad\, \rho \geq 0  \qquad\, \quad\,  \mbox{and}
\; -\rho \leq p \leq\rho \;.
\label{eq:4}
\end{eqnarray}

Thus, using Eq.~\eqref{eq:2} and~\eqref{eq:3} we can easily rewrite the energy conditions as a set of dynamical constraints on scale factor $a(t)$

\begin{eqnarray}
\mbox{\bf NEC}: &&
              \qquad - \frac{\ddot{a}}{a} +
\frac{\dot{a}^2}{a^2}  +
\frac{kc^{2}}{a^2} \geq 0 \;, \nonumber\\
\mbox{\bf WEC}: &&
               - \frac{\ddot{a}}{a} +
\frac{\dot{a}^2}{a^2}  +
\frac{kc^{2}}{a^2} \geq 0\quad\,  \mbox{and} \quad\, \dot{a}^{2}+kc^{2}\geq 0\;, \nonumber\\
\mbox{\bf SEC}: &&
              \qquad\, \qquad\, \frac{\ddot{a}}{a} \leq
0 \;,\nonumber\\
\mbox{\bf DEC}: &&
              \qquad \frac{\ddot{a}}{a}+ 2
              [\frac{\dot{a}^2}{a^2}+\frac{kc^{2}}{a^2}]\geq 0.
\label{eq:5}
\end{eqnarray}

Throughout this paper, we only focus on the flat universe $k=0$. Within this universe, the energy condition of WEC is the same as the energy condition of NEC, so we will neglect the WEC. Mathematically, except $a=0$,  Eq.~\eqref{eq:5} is reasonable for arbitrary real $a(t)$.Hereafter, we use the natural unity, i.e., $\hbar=c=G=k_{B}=1.$

\section{The energy conditions for entropy density}
\label{sec:density}

In classical thermodynamics, the basic quantities are temperature
$T$, heat $Q$, work $W$, internal energy (actually thermal energy or
simply heat) $U$ and entropy $S$. The classical first law is written
as
\begin{equation}
\bigtriangleup U=Q-W.\label{eq:6}
\end{equation}
For an inviscid fluid, the work is given by $dW=pdV$, where $V$ is
the volume and $p$ is the pressure of the fluid so that the first
law can reduce to
\begin{equation}
dU=dQ-pdV.\label{eq:7}
\end{equation}
The classical second law can be read as
\begin{equation}
TdS\geq dQ=dU+dW.\label{eq:8}
\end{equation}
It is clear that energy is an element to characterize the properties
of entropy. Especially, it becomes quite important for an adiabatic
process. Because the energy conditions render very simple and
model-independent bounds associated with energy density $\rho$ and
pressure $p$, it does provide an essential approach that is
instructive to understand the entropy of the universe.

For the perfect fluid, the conservation law of energy-mome\-ntum
tensor reads
\begin{equation}
T^{\mu}_{\phantom{\mu}\nu};_{\mu}=0.\label{eq:9}
\end{equation}
In an expanding universe, it can be written as [26]
\begin{equation}
a^{-3}T\frac{\partial }{\partial t}[\frac{(\rho+p)a^{3}}{T}]=0.\label{eq:10}
\end{equation}
As a result, the entropy density can be defined by
\begin{equation}
s\equiv \frac{\rho+p}{T},\label{eq:11}
\end{equation}
which is proportional to $a^{-3}$, namely
\begin{equation}
s= s_{0}(\frac{a_{0}}{a})^{3},\label{eq:12}
\end{equation}
where $s_{0}$ is a constant, and the subscript $0$ means that the
quantity is evaluated today. Using
\begin{equation}
1+z=\frac{a_{0}}{a},\label{eq:13}
\end{equation}
and combining Eq.~\eqref{eq:12} with~\eqref{eq:13}, we get
\begin{equation}
s=s_{0}(1+z)^{3}.\label{eq:14}
\end{equation}
Since $s_{0}>0$, we easily infer that entropy density $s$ increases
with the redshift $z$. To simplify the discussion, we focus our attention only on the flat universe ($k=0$). Thereby, the energy conditions~\eqref{eq:5} can be rewritten as
\begin{eqnarray}
\textbf{NEC}:&&
    \quad
    s^{-2}[-(\frac{ds}{dt})^{2}+s\frac{d^{2}s}{dt^{2}}]\geq 0, \nonumber\\
\textbf{SEC}:&&
    \quad
    s^{-2}[-4(\frac{ds}{dt})^{2}+3s\frac{d^{2}s}{dt^{2}}]\geq 0, \nonumber\\
\textbf{DEC}:&&
    \quad s^{-2}[2(\frac{ds}{dt})^{2}-s\frac{d^{2}s}{dt^{2}}]\geq 0.
\label{eq:15}
\end{eqnarray}
In fact, in terms of the Hubble parameter
\begin{equation}
H=\frac{\dot{a}}{a}=-\frac{1}{3}s^{-1}\frac{ds}{dt},\label{eq:16}
\end{equation}
the energy conditions can further be expressed as
\begin{eqnarray}
\textbf{NEC}:&&
    \quad H(s)\geq (H_{0}^{2}+\frac{2}{3}ln\frac{s}{s_{0}})^{\frac{1}{2}} ,\nonumber\\
\textbf{SEC}:&&
    \quad H(s)\geq H_{0}(\frac{s}{s_0})^{1/3} ,\nonumber\\
\textbf{DEC}:&&
    \quad H(s)\leq H_{0}\frac{s}{s_0},
\label{eq:17}
\end{eqnarray}
where $H_{0}=\dot{a}(t)/a(t)\big{|}_{z=0}=100\textit{h}$ kms$^{-1}$ Mpc$^{-1}$ is the Hubble parameter today. For a flat
universe, the WMAP 7-year data gives $H_{0}=74.03 \pm 1.42 $ kms$^{-1}$ Mpc$^{-1}$ [39].
\begin{figure}
\begin{center}
\setlength{\floatsep}{8 pt plus 2 pt minus 3 pt}
\includegraphics[width=8.9cm,height=6.0cm]{{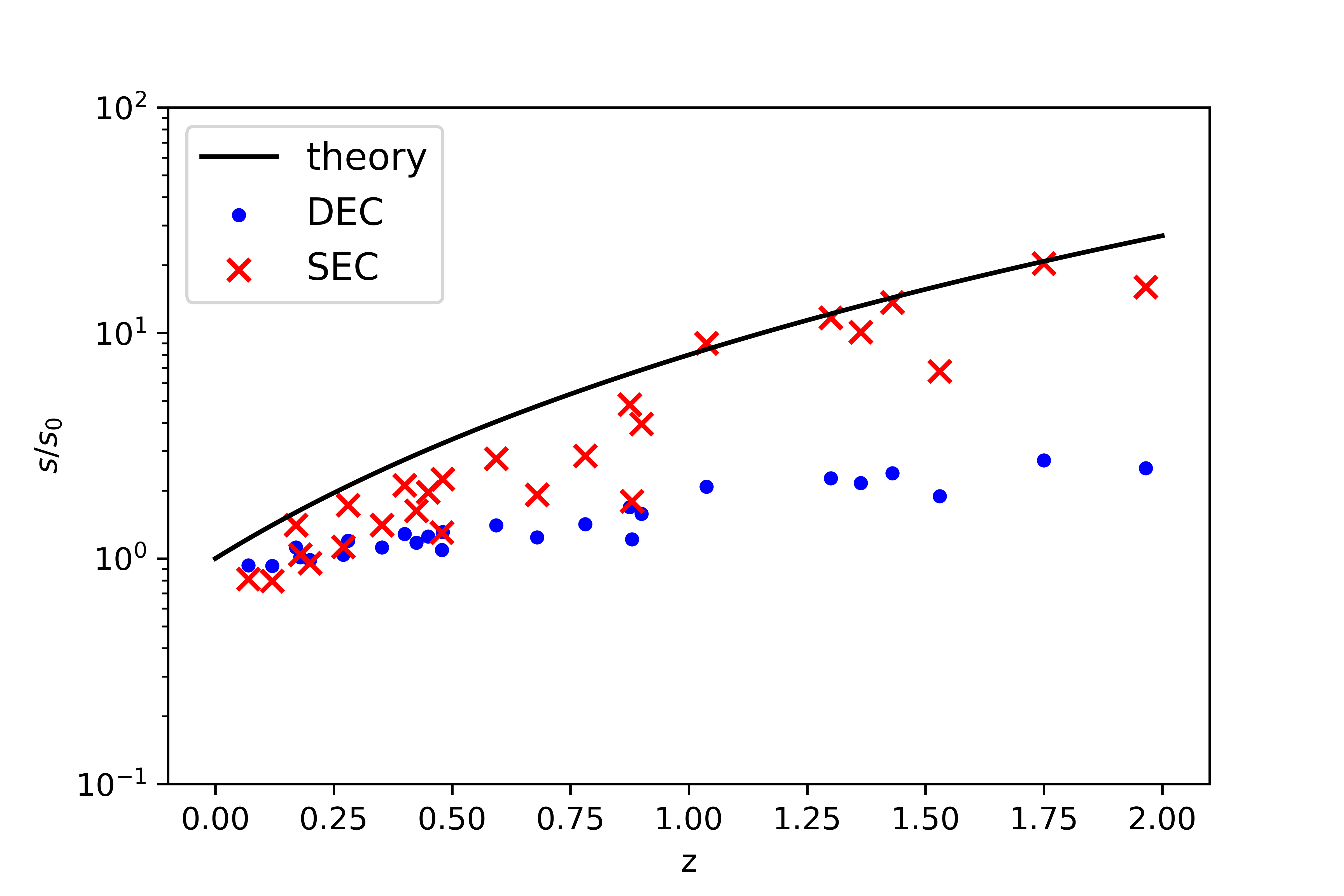}}
\caption{\label{f:zs}  Entropy density as a function of the
redshift. The dots represent the entropy density which is derived
from Eq.~\eqref{eq:18}, using the observational Hubble parameter data (OHD). The data are given in Table 1. The solid curve represents the theoretical prediction, and it is based on Eq.~\eqref{eq:14}. More detailed description is in the main body of the text.}
\end{center}
\end{figure}
\begin{table}
	\centering\caption{The set of available observational $H(z)$ data (OHD)}
	\begin{threeparttable}
	\begin{tabular*}{.48\textwidth}{lccc}
		\hline\hline
		~~~Redshift $z$ &\hspace*{4em} $H(z)\pm 1\sigma$ error \tnote{a} &\hspace*{3em}References \\
		
		\hline
		~~~~~~0.07 &\hspace*{4em} 69 $\pm$ 19.6  &\hspace*{3em} \cite{paper:27} \\
        ~~~~~~0.12 &\hspace*{4em} 68.6 $\pm$ 26.2   & \hspace*{3em}\cite{paper:27} \\
        ~~~~~~0.17 & \hspace*{4em}83 $\pm$ 8  & \hspace*{3em}\cite{paper:41}\\
        ~~~~~~0.1791 & \hspace*{4em}75 $\pm$ 5  & \hspace*{3em}\cite{paper:42} \\
        ~~~~~~0.2  &\hspace*{4em}72.9 $\pm$ 29.6  & \hspace*{3em}\cite{paper:27}\\
        ~~~~~~0.27 &\hspace*{4em} 77 $\pm$ 14  &\hspace*{3em}\cite{paper:41}\\
        ~~~~~~0.28 &\hspace*{4em} 88.8 $\pm$ 36.6  & \hspace*{3em}\cite{paper:27}\\
        ~~~~~~0.3519 &\hspace*{4em}83 $\pm$ 14  & \hspace*{3em}\cite{paper:42}\\
        ~~~~~~0.4  &\hspace*{4em}95 $\pm$ 17 & \hspace*{3em}\cite{paper:41}\\
        ~~~~~~0.4247  &\hspace*{4em}87.1 $\pm$ 11.2 & \hspace*{3em}\cite{paper:43}\\
        ~~~~~~0.4497 &\hspace*{4em}92.8 $\pm$ 12.9 & \hspace*{3em}\cite{paper:43}\\
        ~~~~~~0.4783 &\hspace*{4em}80.9 $\pm$ 9 & \hspace*{3em}\cite{paper:43}\\
        ~~~~~~0.48 &\hspace*{4em} 97 $\pm$ 62 & \hspace*{3em}\cite{paper:44}\\
        ~~~~~~0.5929 &\hspace*{4em} 104 $\pm$ 13 & \hspace*{3em}\cite{paper:42}\\
        ~~~~~~0.6797 &\hspace*{4em} 92 $\pm$ 8 & \hspace*{3em}\cite{paper:42}\\
        ~~~~~~0.7812 &\hspace*{4em} 105 $\pm$ 12 & \hspace*{3em}\cite{paper:42}\\
        ~~~~~~0.8754 &\hspace*{4em} 125 $\pm$ 17 & \hspace*{3em}\cite{paper:42}\\
        ~~~~~~0.88 &\hspace*{4em} 90 $\pm$ 40 & \hspace*{3em}\cite{paper:44}\\
        ~~~~~~0.9 &\hspace*{4em} 117 $\pm$ 23 & \hspace*{3em}\cite{paper:41}\\
        ~~~~~~1.037 &\hspace*{4em} 154 $\pm$ 20 & \hspace*{3em}\cite{paper:42}\\
        ~~~~~~1.3 &\hspace*{4em} 168 $\pm$ 17 & \hspace*{3em}\cite{paper:41}\\
        ~~~~~~1.363 &\hspace*{4em} 160 $\pm$ 33.6 & \hspace*{3em}\cite{paper:45}\\
        ~~~~~~1.43 &\hspace*{4em} 177 $\pm$ 18 & \hspace*{3em}\cite{paper:41}\\
        ~~~~~~1.53 &\hspace*{4em} 140 $\pm$ 14 & \hspace*{3em}\cite{paper:41}\\
        ~~~~~~1.75 &\hspace*{4em} 202 $\pm$ 40 & \hspace*{3em}\cite{paper:41}\\
        ~~~~~~1.965 &\hspace*{4em} 186.5 $\pm$ 50.4 & \hspace*{3em}\cite{paper:45}\\
		\hline\hline
	\end{tabular*}
\begin{tablenotes}
\footnotesize
   \item{a} $H(z)$ figures are in the unit of kms$^{-1}$ Mpc$^{-1}$

   \end{tablenotes}
   	\end{threeparttable}
	\label{tab:comparison}
\end{table}
\subsection{Entropy density}
\label{sec:e&d}

Here, we focus in the constraints on the entropy density. Using Eq.~\eqref{eq:17}, we can obtain
\begin{eqnarray}
\textbf{NEC}:&&
    \quad \frac{s}{s_{0}}\leq e^{\frac{3}{2}(H^{2}-H_{0}^{2})},\nonumber\\
\textbf{SEC}: &&
    \quad \frac{s}{s_{0}}\leq(\frac{H(z)}{H_{0}})^{3},\nonumber\\
\textbf{DEC}: &&
    \quad \frac{s}{s_{0}} \geq \frac{H(z)}{H_{0}}.
    \label{eq:18}
\end{eqnarray}
\begin{figure}
\begin{center}
\includegraphics[width=8.9cm,height=6.0cm]{{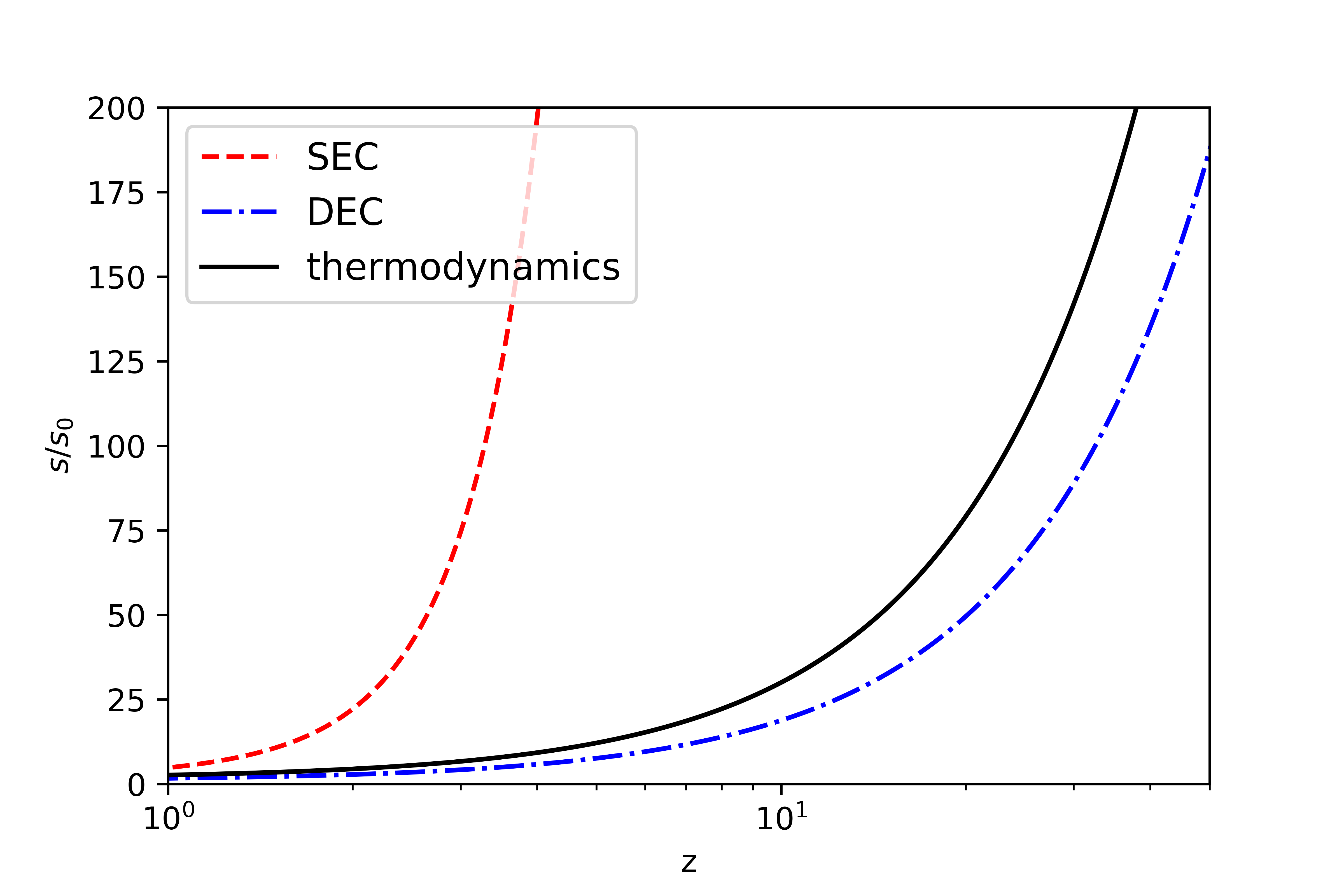}}
\caption{\label{f:H(z)s} Comparison of entropy density $s/s_0$ for two
different cases. The dash-dotted line, dashed line are
speculated by DEC, SEC, respectively. The solid line
corresponds to the first law of thermodynamics (FLT). }
\end{center}
\end{figure}

To get the entropy density with observational data, we use the
observational Hubble parameter $H(z)$ and $H_{0}$, then substitute
them into the Eq.~\eqref{eq:18}.  The observational Hubble parameter $H(z)$ are given in Table 1, and the $H_{0}$ is given in Ref. [39]. In Ref. [27] , $H(z)$ is measured through the redshift $z$ and the derivative of redshift with the respect to cosmic time $dz/dt$ according to Eq.~\eqref{eq:101}.
\begin{eqnarray}
  H(z)=-\frac{1}{1+z}\frac{dz}{dt} \label{eq:101}
\end{eqnarray}
So the observational H(z) data is independent of energy conditions. On the other hand, using the conservation law of energy-momentum tensor,
entropy density is satisfied with the Eq.~\eqref{eq:14}. Therefore, we can compare the theoretical entropy density based on Eq.~\eqref{eq:14} with that from the energy conditions Eq.~\eqref{eq:18} which the observational Hubble parameter is used. As a result, according to the theoretical prediction, using the observational Hubble parameter data (OHD), we can find that DEC is fitter than other energy conditions. In Figure 1, we plot this result.  The dots represent the entropy density which is derived
from Eq.~\eqref{eq:18}, using the observational Hubble parameter data (OHD).
 Clearly, these conditions show that the  entropy
density $s$ depends on $H(z)$, or the dimensionless Hubble
parameter, expansion rate
\begin{eqnarray}
E(z)&=&\frac{H(z)}{H_{0}}\\
&=&\sqrt{\Omega_{m}(1+z)^{3}+\Omega_{R}(1+z)^{4}+\Omega_{\Lambda}+\Omega_{k}(1+z)^{2}}\nonumber
 \label{eq:19}
\end{eqnarray}

\begin{figure}[htbp!!]
    \centering
  \includegraphics[width=8.9cm,height=6.0cm]{{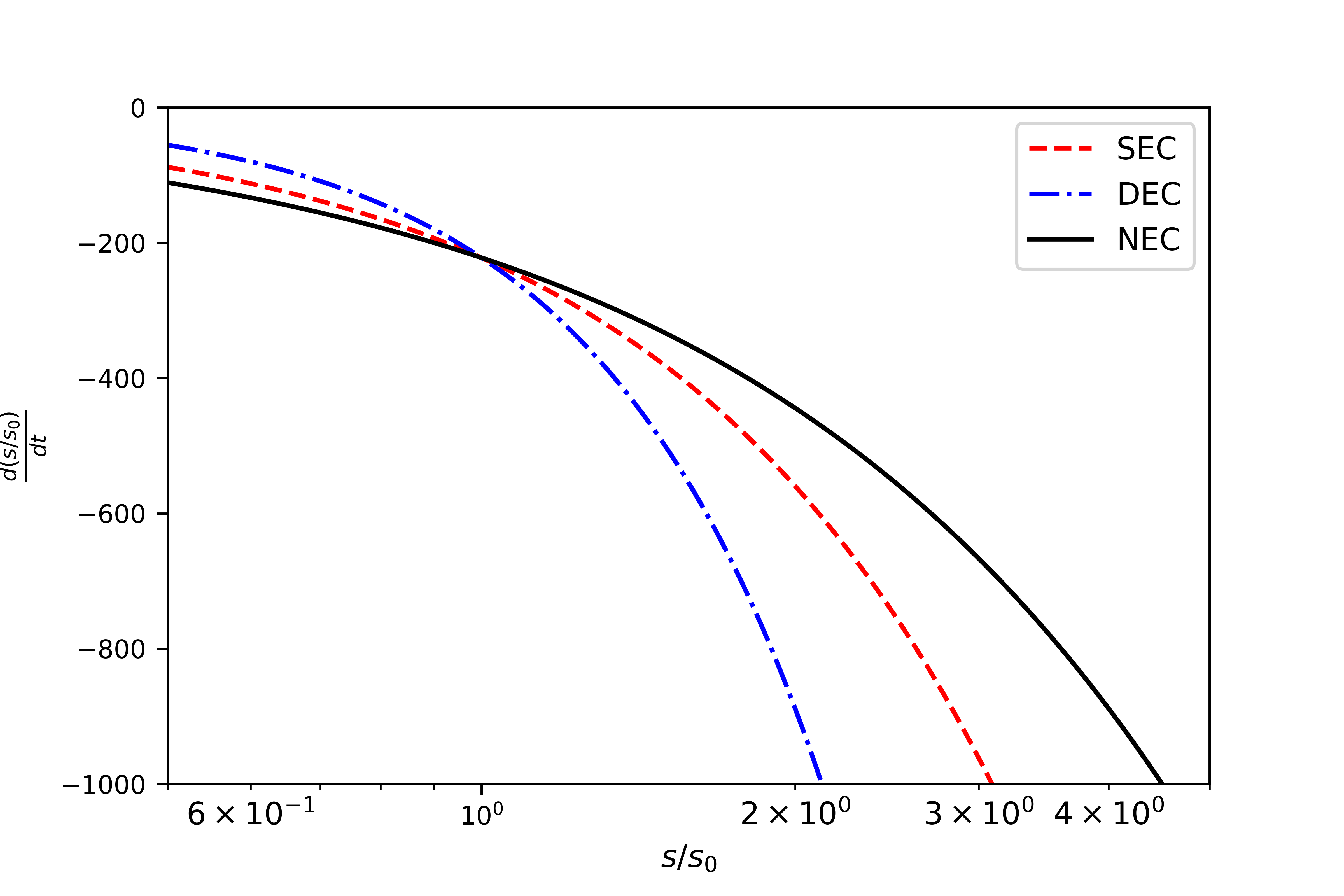}}
    \caption{\label{f:H(z)s}  The $ds/dt$ vs entropy density $s$
       based on Eq.~\eqref{eq:31}. Note that the constraint on $ds/dt$ from NEC and SEC become stronger with the decreasing entropy density $s$.}
\end{figure}

which strongly and explicitly depends upon the four
cosmological parameters, non-relativistic matter energy density
parameter $\Omega_{m}=8\pi G\rho_{m}/(3H^{2}_{0})$, relativistic
matter energy density parameter $\Omega_{R}=8\pi
G\rho_{R}/(3H^{2}_{0})$, vacuum energy density parameter
$\Omega_{\Lambda}=8\pi G\rho_{\Lambda}/(3H^{2}_{0})$, spatial
curvature parameter $\Omega_{k}\equiv -kc^2/(a^{2}_{0}H^{2}_{0})$ and
redshift $z$.

 The cosmological density parameter of the basic $\Lambda$CDM model given
by the WMAP 7-year data [39, 40] are, respectively, the physical
baryon density
 $\Omega_{b}h^{2}=0.02258^{+0.00057}_{-0.00056}$, the
physical cold dark matter density $\Omega_{c}h^{2}=0.1109 \pm0.0056$, and the dark energy density
  $\Omega_{\Lambda}=0.734 \pm0.029$. Therefore, the expansion rate is\\
$E(z)=\sqrt{0.2648(1+z)^{3}+0.734}$. Eventually, we have
\begin{eqnarray}
\textbf{SEC}: &
    \quad \frac{s}{s_0} \leq [0.2648(1+z)^{3}+0.734]^{3/2},\nonumber\\
\textbf{DEC}: &
    \quad \frac{s}{s_0} \geq [0.2648(1+z)^{3}+0.734]^{1/2}.
    \label{eq:20}
\end{eqnarray}

The Eq.~\eqref{eq:20} is illustrated in Figure 2. The dash-dotted line, the dashed line is speculated by DEC, SEC, respectively. According to thermodynamics, we note that the DEC is fitter than other energy conditions,  which is the same as Figure 1. At present, it reaches the same maximum value 1. That is to say, it is not easy to distinguish these conditions now. On the other hand, the universe is expanding all the time. How does the entropy density evolve? Next, we will study the energy conditions from the thermodynamics in Sec.3.2 and compare the change rate of entropy density with the expand rate-Hubble parameter in Sec.3.4.

\subsection{\label{sec:first}The first law of thermodynamics}
Now we turn our attention to the thermodynamics feature of the
entropy density. Before the development of the kinetic theory of
heat, thermodynamics was applied under the assumption that
\begin{figure}[htbp!!]
\centering
\includegraphics[width=8.9cm,height=6.0cm]{{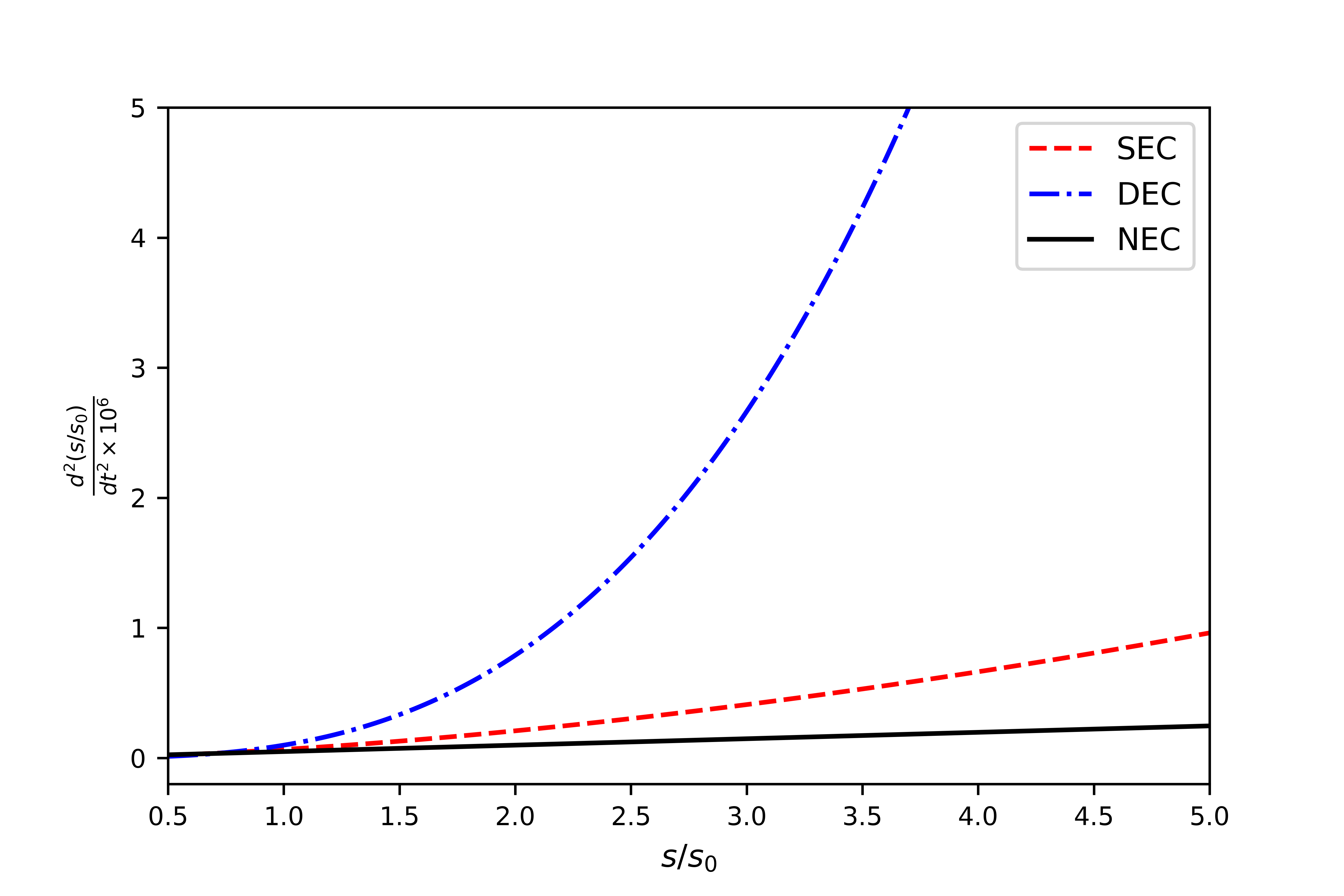}}
      \caption{\label{f:H(z)s} Plot of $d^{2}s/d^{2}t$ as a function of entropy density $s$ from Eq. (31).
      The 2nd derivative of s displays an evolutional trend similar to the 1st derivative. We also note
that it is difficult to present the threshold of $d^{2}s/d^{2}t$ at
DEC.}
\end{figure}
matter is a continuum~\cite{paper:37}. A major step to the understanding of entropy and of the second law of thermodynamics was made following
Boltzmann's statistical interpretation of entropy. It is a notable relationship between entropy and the total number of microstates of a system. Macroscopically, the system is represented
by volume, the number of particles, and given energy. In addition, the
second law of thermodynamics is deemed as an absolute law, i.e.,
entropy always increases in a spontaneous process in an isolated
system-the so-called principle of entropy increase. This is not
different from Newton's laws which are always the obeyed-no exception.
To study the relationship between entropy and the total energy, many
works were made.

In a closed physical system, e.g., a sphere in three spatial
dimensions, the Bekenstein bound~\cite{paper:30}, the ratio of the total
entropy $S$ to total energy $E$ can be expressed as
\begin{equation}
\frac{S}{2\pi RE}\leq 1, \label{eq:21}
\end{equation}
where $R$ denotes the radius of the sphere. This idea has recently
been further elaborated in Ref.~\cite{paper:31,paper:32}.

For an FLRW universe, we take the standard form of the first law of
thermodynamics~\cite{paper:34}
\begin{equation}
TdS=dE+pdV, \label{eq:22}
\end{equation}
where the total entropy $S=sV$, the total energy $E=\rho V$, the
volume of universe $V=\frac{4}{3}\pi\widetilde{r}_{H}^{3}$,
respectively. The Hubble horizon for a flat universe is given by
\begin{equation}
\widetilde{r}_{H}=\frac{c}{H}. \label{eq:23}
\end{equation}
Substituting $S$, $E$, $V$ into Eq.~\eqref{eq:22}, we obtain
\begin{equation}
Td(sV)=d(\rho V)+pdV. \label{eq:24}
\end{equation}
It can be further reduced to
\begin{equation}
(Tds-d\rho)V=(\rho+p-Ts)dV=0, \label{eq:25}
\end{equation}
where the Eq.~\eqref{eq:11} is used. For arbitrary volume, we can obtain
\begin{equation}
Tds-d\rho=0. \label{eq:26}
\end{equation}
If the temperature corresponding to the apparent horizon is defined
as [33]
\begin{equation}
      T=\frac{\hbar c}{2\pi k_{B}\widetilde{r}_{H}}, \label{eq:27}
\end{equation}

Eq.~\eqref{eq:26}can be equivalent to

\begin{equation}
      ds=\frac{2\pi k_{B}}{\hbar c}\frac{c}{H}d\rho. \label{eq:28}
\end{equation}
Substituting Eq.~\eqref{eq:2} into~\eqref{eq:28} , for $k=0$, we have
\begin{equation}
      s=\frac{3}{2}\frac{k_{B}c^{2}}{G\hbar}H. \label{eq:29}
\end{equation}
Consequently, using the natural unity, the Eq.~\eqref{eq:29} can be
simplified to
$$s=\frac{3}{2}H(z).$$

We note that here the entropy density $s$ depends on the Hubble
parameter $H(z)$, which is consistent with the energy conditions. In figure 2, we try to show the comparison between these two kinds of
results, i.e. entropy density from the energy conditions and that
from the first law of thermodynamics. The solid line is obtained
from the first law of thermodynamics. The dotted line, dash-dotted
line and dashed line are obtained from the energy conditions Eq.~\eqref{eq:20}.
Note that the entropy density increases with the increasing
redshift $z$. By the way, if we adopt the above form of $S$, $E$ and
$\widetilde{r}_{H}$, the Bekenstein bound can be read as
\begin{equation}
      s\leq 2\pi c \: \frac{\rho}{H}. \label{eq:30}
\end{equation}

\subsection{\label{sec:deri}Derivatives of entropy}

\label{subsec:deri}
In order to further study the evolution of entropy density, we
should discuss its derivatives from energy conditions. From Eq.~\eqref{eq:15}\eqref{eq:16}\eqref{eq:17},
the energy conditions can be expressed in the form of its
derivatives
\begin{eqnarray}
 \textbf{NEC}:&&
    \frac{d(s/s_0)}{dt}\leq-3\frac{s}{s_0}(H_0^2+\frac{2}{3}ln\frac{s}{s_0})^{1/2} ;\nonumber\\
\textbf{SEC}:&&
    \frac{d(s/s_0)}{dt}\leq-3H_0(\frac{s}{s_0})^{4/3};\nonumber\\
\textbf{DEC}:&&
    \frac{d(s/s_0)}{dt}\geq-3H_0(\frac{s}{s_0})^{2};\nonumber\\
 \textbf{NEC}:&&
    \frac{d^2(s/s_0)}{dt^2}\geq3\frac{s}{s_0}+9\frac{s}{s_0}(H_0^2+\frac{2}{3}ln\frac{s}{s_0});\nonumber\\
\textbf{SEC}:&&
    \frac{d^2(s/s_0)}{dt^2}\geq12H_0^2(\frac{s}{s_0})^{5/3};\nonumber\\
\textbf{DEC}:&&
    \frac{d^2(s/s_0)}{dt^2}\leq18H_0^2(\frac{s}{s_0})^{3}.
    \label{eq:31}
\end{eqnarray}
We find that these derivatives almost all directly depend on entropy
density itself.

The first-order evolution and second-order evolution of entropy
density are presented in Figure 3 and Figure 4, respectively. As described in
Figure 3,  The first-order evolution is negative. Because the entropy
density value is positive, the energy conditions predict that the
entropy density decreases in the expanding universe.

\subsection{The ratio of derivative to Hubble parameter}
In Boltzmann's interpretation of entropy, the correlation of entropy
with disorder is perhaps the earliest, and has its own roots.
An increase in entropy can be associated with an increase in order. That
is to say, entropy characterizes the degree of disorder of a system.
Higher entropy indicates the system is in ``confusion" and
``scatter", while lower entropy indicates the system is in ``order "
 and ``gather". In a free expansion process, particles change from gather to
scatter. The process of work converted into heat energy, is from
order to disorder, i.e., a process of entropy increase. In a local
region, we compare the evolution rate of entropy density with the Hubble
parameter. In terms of Eq.~\eqref{eq:16} and~\eqref{eq:31}, the energy conditions
can be expressed as

\begin{equation}
      \frac{\dot s/s}{\dot a/a}=-3. \label{eq:32}
\end{equation}
It shows that the ratio is a constant, independent of redshift $z$.

\section{Deceleration parameter}
\label{sec:dece}

The cosmic accelerated expansion was primarily inferred from
observations of distant type Ia supernovae (SNe Ia; Riess \textit{et
al.} 1998; Perlmutter \textit{et al.} 1999). It indicates that the
unexpected gravitational physics, which a kind of energy with
negative pressure-so called dark energy, it plays an important role in
the evolution of the universe.

The dark energy model with the Equation of State (constant) $w<-1/3$
makes a positive contribution to the acceleration of the universe,
while the model with $w=-1/3$ has effects on neither the acceleration
nor deceleration. On the other hand, because there still exists
a difficulty in the accurate treatment of the defect of cosmological
models, these models have not been very thoroughly studied (Spergel
$\&$ Pen 1997). Due to this motivation, many works were made to
test these cosmic defect models. For example, the properties of
$\Omega_{DE}-\Omega_{M}$ plane was used to test them in Ref.~\cite{paper:35}.
It is interesting to investigate the behavior of the universe from
the energy conditions. The generalized epoch-dependent deceleration
parameter can be defined as $q(z)=-\ddot{a}/\big(aH^{2}(z)\big)$. As
seen from this definition, the accelerated growth of the cosmic
scale factor $a(t)$ means $\ddot{a}(t) > 0$ which corresponds to
$q(z)<0$, while the decelerated growth of its $\ddot{a}(t)<0$
corresponding to $q(z)>0$. From Eq.~\eqref{eq:12} and~\eqref{eq:16}, we obtain
\begin{equation}
q=3[1+s(\frac{ds}{dt})^{-2}\frac{d^{2}s}{dt^{2}}].\label{eq:33}
\end{equation}
Substituting Eq.~\eqref{eq:15} into Eq.~\eqref{eq:33}, the energy conditions give
\begin{eqnarray}
\textbf{NEC}:&&
    \qquad q(z)\geq 6,\nonumber\\
\textbf{SEC}:&&
    \qquad q(z)\geq 7 ,\nonumber\\
\textbf{DEC}:&&
    \qquad q(z)\leq 9.
    \label{eq:34}
\end{eqnarray}
From Eq.~\eqref{eq:34}, we can obviously obtain $0\leq q(z) \leq2$. In Ref.[4-14], several independent cosmological probes,
such as SN Ia data, CMB, BAO shows that the universe was undergoing acceleration expansion, with Eq.~\eqref{eq:34}, only DEC can fit it.

\section{Conclusions and discussion}
\label{sec:conculsion}
Although some previous works have pointed out some energy conditions
violate the observational data, it did not provide us the detailed information. In this paper, we present an extended analysis
of the energy conditions on entropy density. It is provided that
entropy density increases with the increasing redshift $z$.
According to the theoretical prediction, using the observational Hubble parameter data (OHD), we can find that DEC is fitter than other energy conditions. According to thermodynamics, we note that the DEC is fitter than other energy conditions. Considering the first law of thermodynamics, we have found that $s$ was proportional to Hubble parameter $H(z)$. Generally
speaking, entropy stands for the degree of disorder of the system,
while the Hubble parameter stands for the expansion rate of the universe.
Based on these physics, we also investigate the ratio of the
evolution rate of entropy density to the Hubble parameter of
universe. It shows that the ratio is a constant, independent of
redshift $z$.  At the transition redshift $z_{T}$, the universe
reaches $\ddot{a}(z_{T})=0$ or $q(z_{T})=0$ and evolves from
decelerated to accelerated expansion. In Ref. [5,35], they point out
the transition redshift $z_{T}=0.46 \pm 0.13$.  Several independent cosmological probes show that the universe was undergoing acceleration expansion, only DEC can fit it.

\begin{acknowledgement}
\textbf{Acknowledgements}
We thank Ming-Jian Zhang's preliminary great work on this paper. This work was supported by National Key R\&D Program of China (2017YFA0402600) and the National Science Foundation of China (Grants No. 11929301, 11573006).
\end{acknowledgement}

\end{document}